\documentclass[prl, twocolumn, aps,amsmath,amssymb,amsfonts, superscriptaddress]{revtex4-1}
\usepackage{graphicx} 
\usepackage{graphicx}
\usepackage{graphics}
\usepackage{amsmath}
\usepackage{amsthm} %
\usepackage{amssymb}%
\usepackage{amsfonts}
\usepackage{amssymb}
\usepackage{epstopdf}
\usepackage{makeidx}
\usepackage{epsfig}
\usepackage{amsfonts}
\usepackage{bm}
\usepackage{color}
\usepackage{xcolor}
\usepackage{bbold}
\usepackage{afterpage}
\usepackage{empheq}
\usepackage{hyperref}
\usepackage{booktabs}
\usepackage{subfigure}
\usepackage{makecell}
\usepackage{tikz}
\usepackage{amsmath}
\usetikzlibrary{3d, calc, decorations.markings}
\usetikzlibrary{angles, quotes}
\usepackage{pgfplots}
\usepackage{bm}
\pgfplotsset{compat=1.18}
\newcommand{\be}{\begin{equation}}
\newcommand{\ee}{\end{equation}}
\newcommand{\ba}{\begin{align}}
\newcommand{\ea}{\end{align}}

\begin{document}
\title{Quantum geometry and linear orbital response in arbitrary $SU(2)$ representation}

\author{Rhonald Burgos Atencia}
\affiliation{Dipartimento di Fisica "E. R. Caianiello", Universit\`a di Salerno, IT-84084 Fisciano (SA), Italy}

\begin{abstract}
We develop a unified framework to compute band-geometric quantities in multiband systems whose low-energy Hamiltonians realize arbitrary $SU(2)$ representations. Exploiting the presence of a quantization axis, we use the Wigner--Eckart theorem to identify the allowed interband matrix elements and obtain compact analytic expressions for the quantum geometric tensor, the orbital magnetic moment, and the resulting orbital transport coefficients. The formalism applies to both multifold fermions and gapped $SU(2)$ models. Its versatility is demonstrated through explicit calculations in representative $SU(3)$ and $SU(4)$ settings, where orbital Edelstein and orbital Hall responses arise naturally from the antisymmetric components of the band geometry. Our results reveal a universal link between the algebraic structure of the Hamiltonian and emergent orbitronic phenomena.
\end{abstract}

\maketitle

\section{Introduction}

\textit{Introduction}: Intrinsic contributions to transport phenomena lie at the forefront of modern condensed-matter physics. Within linear-response theory, the anomalous Hall effect stands as the paradigmatic example of a transport coefficient governed by the Berry curvature of the electronic bands \cite{Jungwirth2002,XiaoDi2010,Nagaosa2010,Chang2023}. In recent years, however, growing attention has shifted toward nonlinear responses \cite{GaoYang2014,Sodemann2015,OrtixCarmine2021,Du2021,Du2021II,SuarezRodriguezManuel2025,Jiang2025} where geometric properties of Bloch bands manifest through higher-order multipole moments of the quantum geometric tensor (QGT)\cite{FangYuan2024,GiacomoSala2025}. For instance, at second order in the charge current, the Berry-curvature dipole constitutes the leading geometric contribution in systems preserving time-reversal symmetry \cite{ShaoDingFU2020,Zhao2023,You2018}, whereas in time-reversal–broken systems the dominant response is controlled instead by the dipole of the band-normalized quantum metric \cite{WangChong2021,DasKamal2023,Sharma2025}. Both the Berry curvature and the quantum metric arise as the imaginary and real components, respectively, of the more general QGT. Beyond second order, higher multipole moments of the QGT are expected to appear, reflecting increasingly intricate geometric constraints in multi-band systems and motivating the development of general formalisms capable of capturing these effects.

Another rapidly growing direction in transport theory is orbitronics \cite{Go2021,BurgosAtencia2024}. In this field, the central quantity is no longer the electric charge but the intrinsic orbital magnetic moment (OMM) of Bloch electrons \cite{Sundaram1999,Chang2008}. The OMM is also deeply connected to the underlying band geometry. Indeed, while the Berry curvature can be viewed as arising from virtual interband transitions mediated by the Berry connections, the orbital magnetic moment emerges when these transitions are weighted by the corresponding energy differences. Importantly, this energy weighting implies that even systems with vanishing Berry curvature can host a finite OMM, as in the case of perfectly flat bands. Owing to its geometric origin, the OMM has attracted considerable interest in recent years, particularly because of its central role in orbital transport phenomena such as the orbital Edelstein \cite{CysneTarik2021,CysneTarik2023,LeeJonqiun2024,Gao2025} effect and the orbital Hall effect \cite{Canonico2020,CanonicoLuis2020,Bhowal2021,CysneTarik2021II,CysneTarik2022,PezoArmando2022,Busch2023,PezoArmando2023II,Barbosa2024,CysneTarik2024,LiuHong2024}, key mechanisms underpinning the emerging field of orbitronics.

The role of the OMM has recently expanded into the rapidly developing field of multifold fermions \cite{Robredo2024}. In these systems, the orbital degrees of freedom play a central role in a variety of optical and transport responses, including gyrotropic magnetoelectric effects and circular photogalvanic phenomena \cite{NiZhouliang2020,KaushikSahal2021}. Multifold fermions arise in crystals whose symmetry groups protect band crossings with three-, four-, six-, or eightfold degeneracies at high-symmetry points of the Brillouin zone. Low-energy excitations around these nodes can often be captured by effective Hamiltonians linear in momentum, built from higher-dimensional irreducible representations of angular-momentum operators. In this sense, threefold nodes behave as spin-1 Weyl fermions, while fourfold nodes correspond to spin-3/2 fermions, both exhibiting enhanced geometric and orbital responses compared to conventional Weyl systems. 

Interestingly, the higher-dimensional irreducible representations relevant for multifold fermions obey the SU(2) algebra associated with an effective spin 
$S=(N-1)/2$. This SU(2) structure guarantees the existence of a local quantization axis, along which both the total angular momentum and the magnetic quantum number are well defined and conserved. These properties impose precise selection rules for matrix elements of tensor operators of arbitrary rank, particularly rank-1 and rank-2 operators, corresponding to the Berry connection and the quantum geometric tensor, which together determine the full geometric structure of Bloch states. In this work, we exploit the existence of such a quantization axis to derive general expressions for the quantum geometric tensor and the orbital magnetic moment valid for arbitrary SU(2) representations. A central outcome is that these quantities, as well as their associated orbital responses such as the orbital Hall effect, naturally decompose into a universal geometric factor governed solely by the quantization axis and a multiplet-dependent factor encoding the internal structure of the 
$N$-band manifold, thereby revealing a unified geometric origin for linear and nonlinear orbital transport in multifold systems.

This paper is organized as follows. First we establish the general theoretical framework required to derive the quantum geometric tensor, the orbital magnetic moment, and the intrinsic orbital Hall response for arbitrary SU(2) representations. Second we apply this formalism to the three- and four-dimensional representations relevant for multifold fermions, and we explicitly compute their orbital Edelstein response. Third we illustrate the generality of the quantization-axis construction by considering a massive spin-1 model—widely used to describe low-energy excitations in kagome lattices—and demonstrate how it naturally predicts an orbital Hall effect. 
By “generality of the quantization axis’’ we mean that the axis is not constrained to coincide with the momentum direction: rather, any local direction can serve as a quantization axis as long as the SU(2) algebra is satisfied and the ground-state $I_z$ can be defined. Finally, we summarize our conclusions.

\textit{Theoretical framework}: Our starting point is a Hamiltonian that describes an $N$-dimensional representation of $SU(2)$. 
Such a Hamiltonian can be written as $H = \bm{h} \cdot \bm{I}$,
where the $N$-dimensional matrices $I_i$ satisfy the commutation relations $[I_i, I_j] = i \epsilon_{ijk} I_k$ and $\bm h=\bm h(\bm k)$ is a three dimensional vector. 
This system may or may not be rotationally invariant, but it still allows the definition of a \emph{local quantization axis} along the direction of $\hat{\bm{h}}$. 
The usual choice for the quantization axis in the ground state is the $z$-axis, namely, $I_z |m\rangle = m |m\rangle$.
Using Euler angles, the rotation operator that aligns the ground-state quantization axis with $\hat{\bm{h}}$ is $R(\theta,\phi) = e^{-i\phi I_z} e^{-i\theta I_y}$,
where $\phi = \tan^{-1}(h_y/h_x)$ and $\theta = \cos^{-1}(h_z/|\bm{h}|)$. 
It is straightforward to verify that $R(\theta,\phi)\, I_z\, R^{\dagger}(\theta,\phi) = \hat{\bm{h}}\cdot \bm{I}$.
Accordingly, the eigenstates are $|u_m(\theta,\phi)\rangle = R(\theta,\phi)|m\rangle$,
and the eigenvalues are $\varepsilon_{m\bm k}=m|\bm{h}|$.
These eigenstates define a local frame in momentum space, where the spin (or pseudospin) expectation value is aligned with $\hat{\bm{h}}$. 
The corresponding eigenvalue $m$ thus characterizes the \emph{helicity} of the state, i.e., the projection of the internal angular momentum onto the local momentum direction.
This construction provides the foundation for the evaluation of the system’s topological properties

From the definition of the Berry connection, $\mathcal{A}^{mn}_i = i \langle  u_{m}(\theta,\phi)  | \partial_{k_{i}} | u_{n}(\theta,\phi) \rangle,$
we can define the matrix-valued Berry connection as $\mathcal{A}_{i} = {\bm a}_{i} \cdot \bm I$, with the vector 
${\bm a}_{i}= \left( -\sin\theta\,\frac{\partial \phi}{\partial k_i},
\frac{\partial \theta }{\partial k_i}, \cos\theta\,\frac{\partial \phi}{\partial k_i} \right)$ (See SM).
Fixing a quantization axis is equivalent to choosing a local gauge in the internal (spin or pseudospin) space. 
In our construction, the rotation operator $R(\theta,\phi)$ specifies this gauge, aligning the ground state quantization axis 
with the local direction $\hat{\bm{h}}$. 
As a consequence, variations of $\hat{\bm{h}}$ across momentum space induce a rotation of the local frame, 
and therefore give rise to a gauge connection (the Berry connection) that encodes how the internal degrees of freedom twist under parallel transport. 
The corresponding non-Abelian Berry curvature can then be defined as the covariant derivative of the Berry connection 
$\Omega_{ij} = \partial_i \mathcal{A}_j - \partial_j \mathcal{A}_i - i[\mathcal{A}_i,\mathcal{A}_j]$.
The non-Abelian Berry curvature thus measures the field strength associated with this local frame rotation.
Although the formalism constructs the rotated eigenstates 
$|u_m(\theta,\phi)\rangle = R(\theta,\phi)|m\rangle$, 
the projection onto the original basis $\langle m|\cdots|m\rangle$ is taken when evaluating the Berry curvature. 
Since $|m\rangle$ are eigenstates of $I_z$, the resulting curvature is always proportional to $I_z$. 
This also implies that if the Hamiltonian lacks a component along $I_z$, the projected Berry curvature vanishes.

Since we are interested in calculating the QGT of the system, 
we need to evaluate the product of two Berry connections.
The gauge invariant QGT is defined as $\mathcal{Q}^{mm}_{ij}= \sum_{n\neq m} \mathcal{A}^{mn}_{i}\mathcal{A}^{nm}_{j}$ \cite{Provost1980,YuJiabin2025}. 
The existence of a well-defined local quantization axis ensures that the angular momentum operator $\hat{I}_z$ has quantized eigenvalues. Equivalently, one can always rotate the local basis such that the quantization axis aligns with any chosen local direction, reflecting a form of rotational invariance in the internal degrees of freedom. Once the magnetic quantum number $m$ is well-defined, the allowed transitions between states under tensor operators are dictated by the Wigner–Eckart theorem. In particular, since the Berry connection $\mathcal{A}_i$ transforms as a rank-1 tensor, only transitions between adjacent $m$ states are permitted. 
Then we obtain $\mathcal{Q}^{mm}_{ij}=(\mathcal{A}^{m,m-1}_{i}\mathcal{A}^{m-1,m}_{j} +\mathcal{A}^{m,m+1}_{i}\mathcal{A}^{m+1,m}_{j}).$
Using the ladder operators $I_{\pm} = I_{x}\pm iI_y$ and defining $a^{\pm}_{i} = a^{x}_{i}\pm i a^y_{i}$ we obtain a simple equation for the QGT
\begin{align}
\label{eq:quantumgeometrictensor}
\mathcal{Q}^{mm}_{ij}&=\frac{1}{4} a^{-}_{i} a^{+}_{j} [S(S+1)-m(m-1)] \nonumber \\ 
&+ \frac{1}{4} a^{+}_{i} a^{-}_{j} [S(S+1)-m(m+1)],
\end{align}
with $S$ the total angular momentum. Eq.~\eqref{eq:quantumgeometrictensor} is the first main result of this paper and the starting point to relate the geometry of the band structure to charge and orbital response in arbitrary representations of $SU(2)$ fermions. 

Noting that the real part $\Re(a^{-}_{i} a^{+}_{j})=(a^{x}_{i} a^{x}_{j} +  a^{y}_{i} a^{y}_{j})=\partial_i \hat{\bm n}\cdot \partial_j\hat{\bm n}$
and that the imaginary part $\Im(a^{-}_{i} a^{+}_{j})= (\bm a_i \times \bm a_j)_z =\hat{\bm n}\cdot (\partial_i \hat{\bm n} \times \partial_j\hat{\bm n})$, with $\hat{\bm n}$ a unitary vector in the direction of the
quantization axis, we can calculate the quantum metric and the Berry curvature of the systems. Explicitly we find
the well known result for the Berry curvature 
$\Omega^{mm}_{ij}= -2\Im(\mathcal{Q}^{mm}_{ij})=-m \hat{\bm n}\cdot ( \partial_i\hat{\bm n} \times \partial_j\hat{\bm n})$, which is 
directly related to the magnetic quantum number. The quantum metric reads
\begin{align}
g^{mm}_{ij} &= \Re(\mathcal{Q}^{mm}_{ij})=\frac{1}{2} \partial_i \hat{\bm n}\cdot \partial_j\hat{\bm n} \left[S(S+1)-m^2 \right].
\end{align}
and we note that it depends on both, the magnetic quantum number and on the total angular momentum $S$. 

For the study of responses related to the orbital degree of freedom like orbital Edelstein effect and orbital current, we need the OMM of Bloch electrons.
Applying Wigner-Eckart theorem to its definition 
$\mathcal{M}^{mm}_{l}= \frac{e}{2\hbar}\epsilon_{lij} \Im \sum_{n\neq m}  (\varepsilon_{m,\bm k} - \varepsilon_{n,\bm k} ) \mathcal{A}^{mn}_{i}\mathcal{A}^{nm}_{j}$
one finds the equation (See SM)
\begin{align}
\label{Eq:OMM}
\mathcal{M}^{mm}_{l}&= \frac{e}{8\hbar} \epsilon_{lij} \hat{\bm n}\cdot ( \partial_i\hat{\bm n} \times \partial_j\hat{\bm n} ) B^{(1)}(S,m)
\end{align}
with the definition 
\begin{align}
\label{Eq:factor}
&B^{(\alpha)}(S,m)= 
\Delta_{-}^{\alpha} [S(S+1)-m(m-1)] \Theta(m-1)  \nonumber \\ 
&-
\alpha \Delta_{+}^{\alpha}  [S(S+1)-m(m+1)] \Theta(m+1) 
\end{align}
with $\alpha=1$ and $\Delta^{\alpha}_{\pm}=(\varepsilon_{m,\bm k} - \varepsilon_{m \pm 1,\bm k})^{\alpha}$. The function $\Theta(m\pm 1)=1$ if the state exists and 
$\Theta(m \pm 1)=0$ otherwise. This is relevant because the energies are bounded and hence there are no states beyond/below the highest/lowest
eigenvalue. Hence $m\pm 1 \in \{ -S,... S\}$. We keep this requirement explicit in Eq.~\eqref{Eq:factor} to keep track of the physics, but it is strictly not 
necessary since the factor in squared bracket already takes care of this condition. Eq~\eqref{Eq:OMM} is the second main result in this paper. 
It generalizes previos ones \cite{ZhongShudan2016,Yoda2018} valid only for $S=1/2$ states. 

The orbital Hall effect is the transversal flow of orbital magnetic moment in response to an electric field. In operator 
form, the orbital current is defined as $\mathcal{J}^{\gamma}_{\alpha}=-\frac{\hbar}{g_L\mu_B}\{ \mathcal{M}_{\gamma}, v_{\alpha}\}/2$, where $\mu_B$ is the 
Bohr magneton, $g_L$ is the Lande g-factor and $v_{\alpha}$ is the velocity \cite{Bhowal2021}.
The average current $\langle \mathcal{J}^{\gamma}_{\alpha} \rangle = {\rm tr}(\mathcal{J}^{\gamma}_{\alpha} \rho_{E})$, where $\rho_E$ is the linear response
density matrix \cite{Culcer2017,Burgos2022}. For an electric field $E_{\beta}$ the Hall susceptibility reads (See SM)
\begin{align}
\label{eq:orbitalHall}
\sigma^{\gamma,Hall}_{\alpha\beta}
&=
\frac{e}{g_L\mu_B}
\sum_{\bm k} \hat{\bm n}\cdot ( \partial_{\alpha} \hat{\bm n} \times \partial_{\beta} \hat{\bm n} )
\sum_{m}C_{\gamma}(S,m) 
\end{align}
with
\begin{align}
C(S,m)&= 
\Delta \mathcal{M}^{-}_{\gamma}
[S(S+1)-m(m-1)]n_{FD}(\varepsilon_{m,\bm k}) \Theta_{-} \nonumber  \\ 
&-
\Delta \mathcal{M}^{+}_{\gamma}
[S(S+1)-m(m+1)] n_{FD}(\varepsilon_{m,\bm k}) \Theta_{+}. 
\end{align}
with $\Delta \mathcal{M}^{\pm}_{\gamma}=(\mathcal{M}^{mm}_{\gamma}+\mathcal{M}^{m\pm 1,m \pm 1}_{\gamma})$. Eq.\eqref{eq:orbitalHall} is the 
third main result of this paper and with it the linear orbital response is fully determined for any $N$-dimensional $SU(2)$ representation. 

\begin{table}[t]
\centering
\caption{
Values of $B^{(\alpha)}(S,m)$, defined through the energy differences 
$\Delta^{\alpha}_{\pm} = (\varepsilon_{m}-\varepsilon_{m\pm1})^{\alpha}$. 
For a linear dispersion relation the expressions simplify, since 
$\Delta_{-} = -\Delta_{+} = |\mathbf{h}|$. 
We include $\Theta_{\pm}=\Theta(m\pm 1)$ to keep track of the allowed transitions; 
however, the squared-bracket factor $S(S+1)-m(m\pm1)$ already enforces the 
selection rules, vanishing automatically when the corresponding states lie 
outside the multiplet, i.e., at the boundaries $m=\pm S$.
}
\renewcommand{\arraystretch}{1.25}
\begin{tabular}{c c c c l}
\hline \hline
$S$  & $m$ & $[ \cdots ]\Theta_{-}$ & $[\cdots] \Theta_{+}$ &
$\;B^{(\alpha)}(S,m)$ \\ \hline
\(\tfrac12\) & \(+\tfrac12\)
 & \(1\) & \(0\)
 & \(\;B^{(\alpha)}=(\varepsilon_{1/2}-\varepsilon_{-1/2})^{\alpha}\) \\[4pt]
\(\tfrac12\) & \(-\tfrac12\)
 & \(0\) & \(1\)
 & \(\;B^{(\alpha)}=- \alpha (\varepsilon_{-1/2}-\varepsilon_{1/2})^{\alpha}\) \\[6pt]
\hline \hline
1 & \(+1\)
 & \(2\) & \(0\)
 & \(\;B^{(\alpha)}=2\, (\varepsilon_{1}-\varepsilon_{0})^{\alpha}\) \\[4pt]
1 & \(0\) & \(2\) & \(2\) & $ B^{(\alpha)}  = 2(\varepsilon_{0}-\varepsilon_{-1})^{\alpha}$  \\ [4pt]
 & & & & $ \qquad -2\alpha (\varepsilon_{0}-\varepsilon_{1})^{\alpha}$  \\
1 & \(-1\)
 & \(0\) & \(2\)
 & \(\;B^{(\alpha)}=-2 \alpha (\varepsilon_{-1}-\varepsilon_{0})^{\alpha}\) \\[6pt]
\hline \hline
\(\tfrac32\) & \(+\tfrac32\)
 & \(3\) & \(0\)
 & \(\;B^{(\alpha)}=3  (\varepsilon_{3/2}-\varepsilon_{1/2})^{\alpha}\) \\[4pt]
\(\tfrac32\) & \(+\tfrac12\)
 & \(4\) & \(3\)
 & \(\;B^{(\alpha)}=4\,  (\varepsilon_{1/2}-\varepsilon_{-1/2}) ^{\alpha}  \) \\[4pt]
 &  &  &  & \qquad \(\ - 3 \alpha \, (\varepsilon_{1/2}-\varepsilon_{3/2})  ^{\alpha} \) \\[4pt]
\(\tfrac32\) & \(-\tfrac12\)
 & \(3\) & \(4\)
 & \(\;B^{(\alpha)}=3\, (\varepsilon_{-1/2}-\varepsilon_{-3/2})^{\alpha}  \) \\[4pt]
&&& & \(\; \qquad  -4 \alpha \, (\varepsilon_{-1/2}-\varepsilon_{1/2}) ^{\alpha}\) \\[4pt]
\(\tfrac32\) & \(-\tfrac32\)
 & \(0\) & \(3\)
 & \(\;B^{(\alpha)}=-3 \alpha (\varepsilon_{-3/2}-\varepsilon_{-1/2})^{\alpha}\)  \\
\hline \hline
\end{tabular}
\label{Tab:generalterms}
\end{table}

\textit{Multifold fermions}: $N$-dimensional representations of $SU(2)$ arise, for instance, in multifold fermions. Near symmetry-protected band-crossing points, the Hamiltonian takes the general form considered here. For multifold fermions, additional simplifications occur because the quantization axis aligns with $\hat{\bm{k}}$, leading to $\hat{\bm{n}}\cdot(\partial_i \hat{\bm{n}} \times \partial_j \hat{\bm{n}})= \epsilon_{ijl}k_l/k^{3}$. A three-dimensional representation of $SU(2)$ is realized by 
the Hamiltonian \cite{Flicker2018} 
\begin{equation}
H_{3f}=
\begin{pmatrix}
 0     & e^{i\phi}k_x & e^{-i\phi}k_y  \\
e^{-i\phi}k_x      &  0 & e^{i\phi}k_z \\
e^{i\phi}k_y     & e^{-i\phi}k_z & 0
\end{pmatrix}
\end{equation}
where $\phi$ is a material-dependent parameter (not to be confused with the azimuthal angle on the Bloch sphere). For $\phi=(2n+1)\pi/6$, this Hamiltonian reduces exactly to the spin-1 representation of $SU(2)$. A four-dimensional representation of $SU(2)$ can also be realized through the effective Hamiltonian \cite{Flicker2018}
\begin{equation}
H_{4f}=
\begin{pmatrix}
a k_z & 0 &
-\frac{(a+3b)k^{+}}{4} &
\frac{\sqrt{3}(a-b)k^{-}}{4}  \\
0 & b k_z &
\frac{\sqrt{3}(a-b)k^{-}}{4}  &
-\frac{(3a+b)k^{+}}{4}  \\
-\frac{(a+3b)k^{-}}{4}  &
\frac{\sqrt{3}(a-b)k^{+}}{4}  &
- a k_z & 0 \\
\frac{\sqrt{3}(a-b)k^{+}}{4} &
-\frac{(3a+b)k^{-}}{4}&
0 & - b k_z
\end{pmatrix},
\end{equation}
with $k^\pm = k_x \pm i k_y$ and $a,b$ material dependent parameters. This model describes fermionic states that satisfy the commutation relations $[I_i, I_j] = i\lambda\, \epsilon_{ijl} I_l$ with a single scalar $\lambda$, provided the parameters fulfill $(a-b)(a+3b)(3a+b)=0$. In that case, the coefficient $\lambda = \frac{(a+b)a^{2}-10ab+b^{2}}{4(a^{2}+b^{2})}$ acts as an effective $\hbar$ for the emergent $SU(2)$ algebra (see SM). Although the case $a=b$ is mathematically allowed, it represents two copies of spin-$1/2$ (SM). 
However, $a=-b/3$ or $a=-3b$ leads to four dimensional representations of $SU(2)$.

The orbital Edelstein effect is the orbital magnetization due to an external electric field. 
It is defined as $M_E = {\rm tr}\!\left(\mathcal{M}^{mm}_{i}\,\rho^{mm}_{E}\right)$, where
$\rho^{mm}_E$ is the linear response band diagonal density matrix (See SM) \cite{Culcer2017,Burgos2022}. 
In linear response and at zero temperature, this definition leads to the susceptibility (see SM for details).
\begin{align}
\alpha_{ij} &
= -\tau \frac{e^2}{8\hbar} \sum_{\bm k} \epsilon_{ils} \hat{\bm n}\cdot ( \partial_l \hat{\bm n} \times \partial_s \hat{\bm n} ) \nonumber \\
&\times \sum_{m}   B^{(1)}(S,m) \frac{1}{\hbar} \frac{\partial\varepsilon_{m\bm k}  }{\partial k_j} 
\delta( \varepsilon_{m\bm k}- \varepsilon_F ).
\end{align}
with $\tau$ a relaxation time. From table~\eqref{Tab:generalterms} we extract the OMM for the spin-$1$ and spin-$3/2$ cases. For spin $S=1$ we obtain
\begin{align}
\alpha^{S=1}_{zz} &=
-\tau \frac{e^2 v_F}{8\hbar}\,\frac{2}{3}\,\frac{1}{(2\pi)^2}
\begin{cases}
2\varepsilon_F, &  m=1 \\[2pt]
4\varepsilon_F, (=0) & m=0 \\[2pt]
2\varepsilon_F, & m=-1,
\end{cases}
\end{align}
where the flat band yields a vanishing contribution because it does not form a Fermi surface. However, a small quadratic correction to the Hamiltonian lifts this cancellation and produces a finite response~\cite{Flicker2018}. For $S=3/2$ we find
\begin{align}
\alpha^{S=3/2}_{zz} &=
-\tau \frac{e^2 v_F}{8\hbar}\,\frac{2}{3}\,\frac{1}{(2\pi)^2}
\begin{cases}
3\varepsilon_F,   &  m=3/2 \\[2pt]
7\varepsilon_F,   &  m=1/2 \\[2pt]
7\varepsilon_F,   &  m=-1/2 \\[2pt]
3\varepsilon_F,   &  m=-3/2.
\end{cases}
\end{align}

These expressions for the orbital magnetization are consistent with the gyrotropic magneto-electric effect reported in Ref.~\cite{Flicker2018}, up to an overall constant factor, since both 
responses relies on the OMM at the Fermi surface. The trend is transparent: increasing the spin dimensionality enhances the response, although the final value depends sensitively on which bands are intersected by the Fermi level. For instance, if $\varepsilon_F$ crosses both the $m=3/2$ and $m=1/2$ branches, the total spin-$3/2$ response scales as $10\varepsilon_F$.

\begin{figure}[t] 
\centering
\includegraphics[width=1\columnwidth]{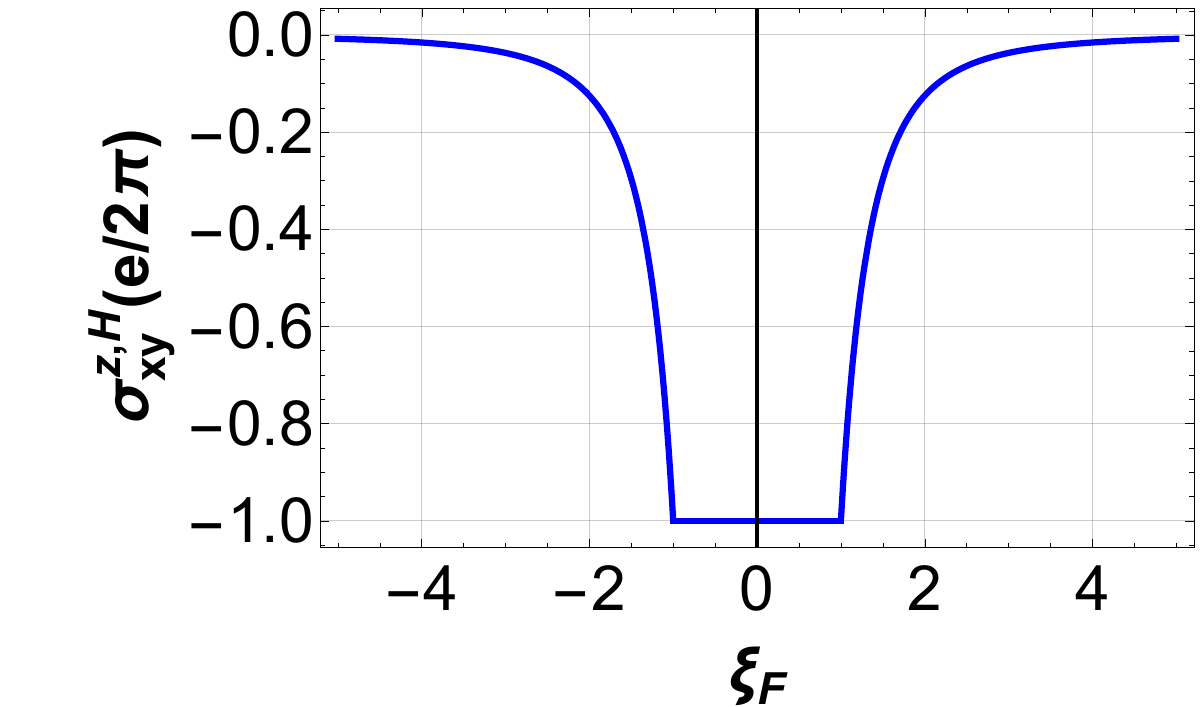} \\
\includegraphics[width=1\columnwidth]{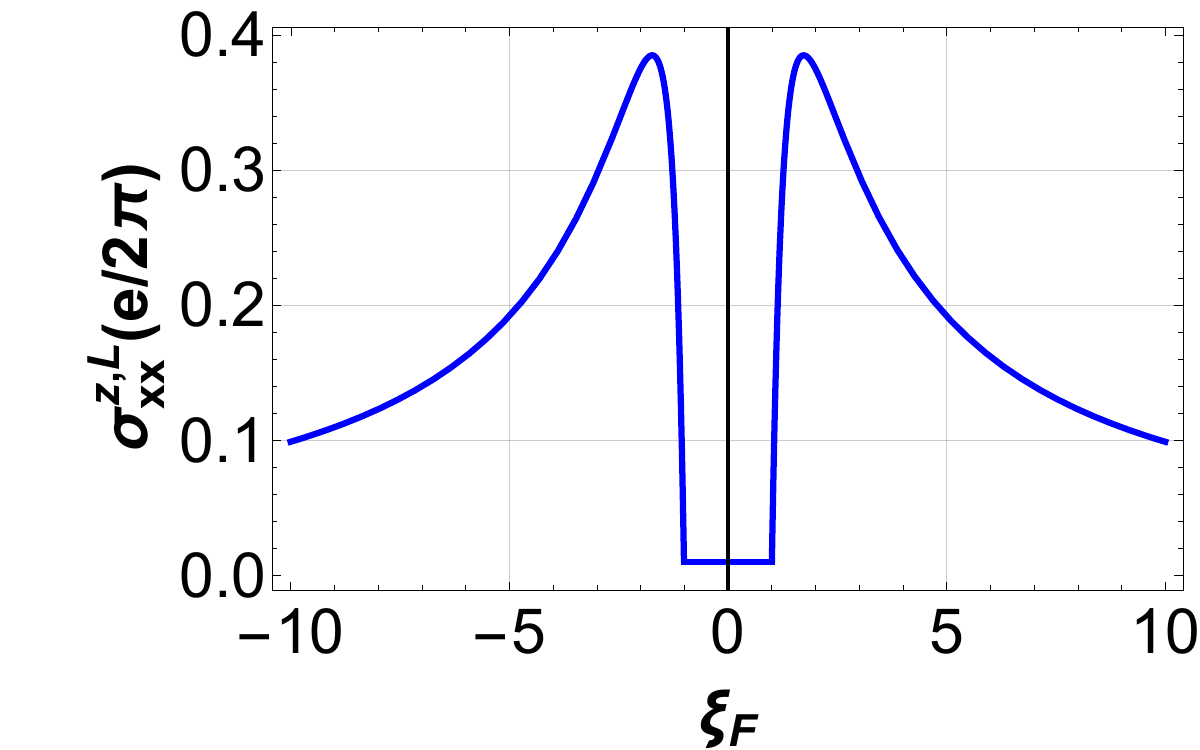} 
\caption{ 
Top panel: orbital Hall response. Bottom panel: longitudinal orbital response. 
Both quantities are plotted as functions of the dimensionless ratio 
$\xi_F=\varepsilon_F/M$ and expressed in units of $e/2\pi$. 
The magnitude of the orbital Hall signal is controlled by the prefactor 
$C_H=(e/g_L\mu_B)(\hbar v_F^2/M)$, which has been factored out of the curves to 
highlight the universal dependence on $\xi_F$. 
Similarly, the longitudinal response carries the prefactor 
$C_L=(e\,\tau\, v_F^2)/(8 g_L \mu_B)$, also omitted from the plots. 
Using realistic parameters for Dirac-like quasiparticles reported in the literature \cite{YeLinda2018,KangMinguII2020,KangMingu2020}
($v_F \approx 10^5\,\mathrm{m/s}$, $\tau \approx 10\,\mathrm{ps}$, 
$M \approx 10\,\mathrm{meV}$, and $g_L=2$), we estimate $C_H \approx 5$ and 
$C_L \approx 10$. These values illustrate that both orbital Hall and longitudinal 
orbital responses can reach experimentally relevant magnitudes in systems with 
moderate band velocities and small but finite gaps.
 }
\label{fig:orbital}
\end{figure}

\textit{Massive spin-$1$ states}. The formalism developed above only requires the existence of a well-defined quantization axis, independent of its orientation. 
Exploiting this property, we predict a linear longitudinal orbital current and an orbital Hall current in massive spin-$1$ systems. Such states arise as low-energy excitations in kagome lattices and are described by the Hamiltonian $H=\bm{d}\cdot\bm{I}$, with $\bm{d}=(k_x,k_y,M)$ and $\bm{I}$ the three-dimensional representation of the $SU(2)$ fermions, where $M$ is half the gap \cite{GreenDmitry2010}. The orbital Hall response follows as (See SM)
\begin{align}
\sigma^{z,Hall}_{xy} 
&= -\frac{e}{2\pi} \left( \frac{e}{g_L \mu_B} \frac{\hbar v_F^2}{M} \right) \frac{1}{\xi^3_F}  \quad \textit{for} \quad \varepsilon_F > M
\end{align}
with $\xi_F=\varepsilon_F/M$. We plot the orbital Hall response in Fig.~\eqref{fig:orbital}. As expected it has its maximum values close to the gap, namely, band geometric contributions are expected to be stronger close to the band touching point. 

We derive a longitudinal orbital current in the supplementary material. After some algebra the result reads
\begin{align}
\sigma^{z,(L)}_{xx}
&= \frac{e}{2\pi} \left(\frac{e\tau v^2_F}{8g_L\mu_B}  \right)
\frac{ (\xi^2_F -1) }{\xi^3_F}  \quad \varepsilon_F > M.
\end{align} 

We plot the longitudinal response in Fig.~\eqref{fig:orbital} as well. As a Fermi surface response, it vanishes at
the bottom/top of the conduction/valence band. It also develops a peak, characteristic of Berry curvature driven 
responses, which also characterizes inter-band transitions effects. 
The orbital Hall angle $\beta_{OHA}=\sigma^{z,Hall}_{xy}/\sigma^{z,(L)}_{xx} \propto \hbar /\tau M (\xi^2_F -1)$ indicates the dominance of the Hall response close to the gap. 

\textit{Discussion}---We close with a few remarks on the band-normalized quantum metric for $n$-fold fermions. It takes the form 
$\mathcal{G}^{mm}_{ij}=\frac{1}{4}\,\partial_i\hat{\bm n}\!\cdot\!\partial_j\hat{\bm n}\, B^{(-1)}(S,m)$, 
with the values of $B^{(\alpha)}(S,m)$ for $\alpha=-1$ listed in Table~\ref{Tab:generalterms}. 
For multifold fermions, the quantization axis aligns with $\hat{\bm k}$, so the geometric factor simplifies to  
$\partial_i\hat{\bm n}\cdot\partial_j\hat{\bm n} = (\delta_{ij}-k_i k_j/k^2)/k^2$.  
With this expression, second-order responses can be computed whenever a mechanism breaking inversion symmetry is present.

The present formalism naturally extends to include in-plane, non-quantizing (Zeeman-type) magnetic fields, enabling the study of planar Hall responses and magnetoresistance in higher-spin systems. Such fields leave the underlying $SU(2)$ algebra intact and therefore preserve both the existence of the quantization axis and the associated selection rules.  
Regarding the orbital Hall effect in multifold fermions, its presence is indeed expected; however, continuum models become unreliable due to the $1/k$ behavior of the integrand, which forces the introduction of an artificial momentum cutoff.  
In contrast, tight-binding realizations of $SU(2)$ fermions naturally confine momenta to the Brillouin zone, thereby regularizing the integrals and eliminating spurious logarithmic divergences~\cite{QunYang2023}.

A central outcome of our work is that all geometric quantities discussed here—including the QGT, the OMM, and the resulting orbital responses—exhibit a natural decomposition into (i) a universal geometric factor determined solely by the orientation and texture of the quantization axis, and (ii) a multiplet-dependent factor encoding the internal structure of the $N$-band manifold. This decomposition exposes a common geometric origin underlying linear and nonlinear orbital responses across a wide class of $SU(2)$-based systems.

\textit{Conclusions}---We developed a unified framework to compute the quantum geometric tensor, the orbital magnetic moment, and both longitudinal and Hall orbital currents for arbitrary representations of the $SU(2)$ algebra. The approach relies only on the existence of a local quantization axis—guaranteed by the $SU(2)$ structure—and on the selection rules imposed by the Wigner--Eckart theorem.  
Explicit calculations for three- and four-dimensional representations show that the resulting orbital magnetization agrees with known mechanisms such as the gyrotropic magneto-electric effect.  
Moreover, by analyzing a massive spin-1 model, we predict a finite orbital Hall response in gapped systems, together with a longitudinal orbital current.  
Our results provide a systematic route for understanding and predicting orbital transport phenomena in materials hosting higher-spin or multifold fermions, and open the way to generalizing orbitronic responses in broad classes of multiband systems with nontrivial algebraic structure.

\begin{acknowledgments}
We acknowledge support by the Italian Ministry
of Foreign Affairs and International Cooperation, grant
PGR12351 “ULTRAQMAT” and the PNRR MUR project PE0000023-NQSTI.).
\end{acknowledgments}

\bibliography{THG}

\begin{thebibliography}{53}%
\makeatletter
\providecommand \@ifxundefined [1]{%
 \@ifx{#1\undefined}
}%
\providecommand \@ifnum [1]{%
 \ifnum #1\expandafter \@firstoftwo
 \else \expandafter \@secondoftwo
 \fi
}%
\providecommand \@ifx [1]{%
 \ifx #1\expandafter \@firstoftwo
 \else \expandafter \@secondoftwo
 \fi
}%
\providecommand \natexlab [1]{#1}%
\providecommand \enquote  [1]{``#1''}%
\providecommand \bibnamefont  [1]{#1}%
\providecommand \bibfnamefont [1]{#1}%
\providecommand \citenamefont [1]{#1}%
\providecommand \href@noop [0]{\@secondoftwo}%
\providecommand \href [0]{\begingroup \@sanitize@url \@href}%
\providecommand \@href[1]{\@@startlink{#1}\@@href}%
\providecommand \@@href[1]{\endgroup#1\@@endlink}%
\providecommand \@sanitize@url [0]{\catcode `\\12\catcode `\$12\catcode
  `\&12\catcode `\#12\catcode `\^12\catcode `\_12\catcode `\%12\relax}%
\providecommand \@@startlink[1]{}%
\providecommand \@@endlink[0]{}%
\providecommand \url  [0]{\begingroup\@sanitize@url \@url }%
\providecommand \@url [1]{\endgroup\@href {#1}{\urlprefix }}%
\providecommand \urlprefix  [0]{URL }%
\providecommand \Eprint [0]{\href }%
\providecommand \doibase [0]{http://dx.doi.org/}%
\providecommand \selectlanguage [0]{\@gobble}%
\providecommand \bibinfo  [0]{\@secondoftwo}%
\providecommand \bibfield  [0]{\@secondoftwo}%
\providecommand \translation [1]{[#1]}%
\providecommand \BibitemOpen [0]{}%
\providecommand \bibitemStop [0]{}%
\providecommand \bibitemNoStop [0]{.\EOS\space}%
\providecommand \EOS [0]{\spacefactor3000\relax}%
\providecommand \BibitemShut  [1]{\csname bibitem#1\endcsname}%
\let\auto@bib@innerbib\@empty
\bibitem [{\citenamefont {Jungwirth}\ \emph {et~al.}(2002)\citenamefont
  {Jungwirth}, \citenamefont {Niu},\ and\ \citenamefont
  {MacDonald}}]{Jungwirth2002}%
  \BibitemOpen
  \bibfield  {author} {\bibinfo {author} {\bibfnamefont {T.}~\bibnamefont
  {Jungwirth}}, \bibinfo {author} {\bibfnamefont {Q.}~\bibnamefont {Niu}}, \
  and\ \bibinfo {author} {\bibfnamefont {A.~H.}\ \bibnamefont {MacDonald}},\
  }\href {\doibase 10.1103/PhysRevLett.88.207208} {\bibfield  {journal}
  {\bibinfo  {journal} {Phys. Rev. Lett.}\ }\textbf {\bibinfo {volume} {88}},\
  \bibinfo {pages} {207208} (\bibinfo {year} {2002})}\BibitemShut {NoStop}%
\bibitem [{\citenamefont {Xiao}\ \emph {et~al.}(2010)\citenamefont {Xiao},
  \citenamefont {Chang},\ and\ \citenamefont {Niu}}]{XiaoDi2010}%
  \BibitemOpen
  \bibfield  {author} {\bibinfo {author} {\bibfnamefont {D.}~\bibnamefont
  {Xiao}}, \bibinfo {author} {\bibfnamefont {M.-C.}\ \bibnamefont {Chang}}, \
  and\ \bibinfo {author} {\bibfnamefont {Q.}~\bibnamefont {Niu}},\ }\href
  {\doibase 10.1103/RevModPhys.82.1959} {\bibfield  {journal} {\bibinfo
  {journal} {Rev. Mod. Phys.}\ }\textbf {\bibinfo {volume} {82}},\ \bibinfo
  {pages} {1959} (\bibinfo {year} {2010})}\BibitemShut {NoStop}%
\bibitem [{\citenamefont {Nagaosa}\ \emph {et~al.}(2010)\citenamefont
  {Nagaosa}, \citenamefont {Sinova}, \citenamefont {Onoda}, \citenamefont
  {MacDonald},\ and\ \citenamefont {Ong}}]{Nagaosa2010}%
  \BibitemOpen
  \bibfield  {author} {\bibinfo {author} {\bibfnamefont {N.}~\bibnamefont
  {Nagaosa}}, \bibinfo {author} {\bibfnamefont {J.}~\bibnamefont {Sinova}},
  \bibinfo {author} {\bibfnamefont {S.}~\bibnamefont {Onoda}}, \bibinfo
  {author} {\bibfnamefont {A.~H.}\ \bibnamefont {MacDonald}}, \ and\ \bibinfo
  {author} {\bibfnamefont {N.~P.}\ \bibnamefont {Ong}},\ }\href {\doibase
  10.1103/RevModPhys.82.1539} {\bibfield  {journal} {\bibinfo  {journal} {Rev.
  Mod. Phys.}\ }\textbf {\bibinfo {volume} {82}},\ \bibinfo {pages} {1539}
  (\bibinfo {year} {2010})}\BibitemShut {NoStop}%
\bibitem [{\citenamefont {Chang}\ \emph {et~al.}(2023)\citenamefont {Chang},
  \citenamefont {Liu},\ and\ \citenamefont {MacDonald}}]{Chang2023}%
  \BibitemOpen
  \bibfield  {author} {\bibinfo {author} {\bibfnamefont {C.-Z.}\ \bibnamefont
  {Chang}}, \bibinfo {author} {\bibfnamefont {C.-X.}\ \bibnamefont {Liu}}, \
  and\ \bibinfo {author} {\bibfnamefont {A.~H.}\ \bibnamefont {MacDonald}},\
  }\href {\doibase 10.1103/RevModPhys.95.011002} {\bibfield  {journal}
  {\bibinfo  {journal} {Rev. Mod. Phys.}\ }\textbf {\bibinfo {volume} {95}},\
  \bibinfo {pages} {011002} (\bibinfo {year} {2023})}\BibitemShut {NoStop}%
\bibitem [{\citenamefont {Gao}\ \emph {et~al.}(2014)\citenamefont {Gao},
  \citenamefont {Yang},\ and\ \citenamefont {Niu}}]{GaoYang2014}%
  \BibitemOpen
  \bibfield  {author} {\bibinfo {author} {\bibfnamefont {Y.}~\bibnamefont
  {Gao}}, \bibinfo {author} {\bibfnamefont {S.~A.}\ \bibnamefont {Yang}}, \
  and\ \bibinfo {author} {\bibfnamefont {Q.}~\bibnamefont {Niu}},\ }\href
  {\doibase 10.1103/PhysRevLett.112.166601} {\bibfield  {journal} {\bibinfo
  {journal} {Phys. Rev. Lett.}\ }\textbf {\bibinfo {volume} {112}},\ \bibinfo
  {pages} {166601} (\bibinfo {year} {2014})}\BibitemShut {NoStop}%
\bibitem [{\citenamefont {Sodemann}\ and\ \citenamefont
  {Fu}(2015)}]{Sodemann2015}%
  \BibitemOpen
  \bibfield  {author} {\bibinfo {author} {\bibfnamefont {I.}~\bibnamefont
  {Sodemann}}\ and\ \bibinfo {author} {\bibfnamefont {L.}~\bibnamefont {Fu}},\
  }\href {\doibase 10.1103/PhysRevLett.115.216806} {\bibfield  {journal}
  {\bibinfo  {journal} {Phys. Rev. Lett.}\ }\textbf {\bibinfo {volume} {115}},\
  \bibinfo {pages} {216806} (\bibinfo {year} {2015})}\BibitemShut {NoStop}%
\bibitem [{\citenamefont {Ortix}(2021)}]{OrtixCarmine2021}%
  \BibitemOpen
  \bibfield  {author} {\bibinfo {author} {\bibfnamefont {C.}~\bibnamefont
  {Ortix}},\ }\href {\doibase https://doi.org/10.1002/qute.202100056}
  {\bibfield  {journal} {\bibinfo  {journal} {Advanced Quantum Technologies}\
  }\textbf {\bibinfo {volume} {4}},\ \bibinfo {pages} {2100056} (\bibinfo
  {year} {2021})},\ \Eprint
  {http://arxiv.org/abs/https://advanced.onlinelibrary.wiley.com/doi/pdf/10.1002/qute.202100056}
  {https://advanced.onlinelibrary.wiley.com/doi/pdf/10.1002/qute.202100056}
  \BibitemShut {NoStop}%
\bibitem [{\citenamefont {Du}\ \emph {et~al.}(2021{\natexlab{a}})\citenamefont
  {Du}, \citenamefont {Wang}, \citenamefont {Sun}, \citenamefont {Lu},\ and\
  \citenamefont {Xie}}]{Du2021}%
  \BibitemOpen
  \bibfield  {author} {\bibinfo {author} {\bibfnamefont {Z.~Z.}\ \bibnamefont
  {Du}}, \bibinfo {author} {\bibfnamefont {C.~M.}\ \bibnamefont {Wang}},
  \bibinfo {author} {\bibfnamefont {H.-P.}\ \bibnamefont {Sun}}, \bibinfo
  {author} {\bibfnamefont {H.-Z.}\ \bibnamefont {Lu}}, \ and\ \bibinfo {author}
  {\bibfnamefont {X.~C.}\ \bibnamefont {Xie}},\ }\href {\doibase
  10.1038/s41467-021-25273-4} {\bibfield  {journal} {\bibinfo  {journal}
  {Nature Communications}\ }\textbf {\bibinfo {volume} {12}},\ \bibinfo {pages}
  {5038} (\bibinfo {year} {2021}{\natexlab{a}})}\BibitemShut {NoStop}%
\bibitem [{\citenamefont {Du}\ \emph {et~al.}(2021{\natexlab{b}})\citenamefont
  {Du}, \citenamefont {Lu},\ and\ \citenamefont {Xie}}]{Du2021II}%
  \BibitemOpen
  \bibfield  {author} {\bibinfo {author} {\bibfnamefont {Z.~Z.}\ \bibnamefont
  {Du}}, \bibinfo {author} {\bibfnamefont {H.-Z.}\ \bibnamefont {Lu}}, \ and\
  \bibinfo {author} {\bibfnamefont {X.~C.}\ \bibnamefont {Xie}},\ }\href
  {\doibase 10.1038/s42254-021-00359-6} {\bibfield  {journal} {\bibinfo
  {journal} {Nature Reviews Physics}\ }\textbf {\bibinfo {volume} {3}},\
  \bibinfo {pages} {744} (\bibinfo {year} {2021}{\natexlab{b}})}\BibitemShut
  {NoStop}%
\bibitem [{\citenamefont {Su{\'a}rez-Rodr{\'\i}guez}\ \emph
  {et~al.}(2025)\citenamefont {Su{\'a}rez-Rodr{\'\i}guez}, \citenamefont
  {de~Juan}, \citenamefont {Souza}, \citenamefont {Gobbi}, \citenamefont
  {Casanova},\ and\ \citenamefont {Hueso}}]{SuarezRodriguezManuel2025}%
  \BibitemOpen
  \bibfield  {author} {\bibinfo {author} {\bibfnamefont {M.}~\bibnamefont
  {Su{\'a}rez-Rodr{\'\i}guez}}, \bibinfo {author} {\bibfnamefont
  {F.}~\bibnamefont {de~Juan}}, \bibinfo {author} {\bibfnamefont
  {I.}~\bibnamefont {Souza}}, \bibinfo {author} {\bibfnamefont
  {M.}~\bibnamefont {Gobbi}}, \bibinfo {author} {\bibfnamefont
  {F.}~\bibnamefont {Casanova}}, \ and\ \bibinfo {author} {\bibfnamefont
  {L.~E.}\ \bibnamefont {Hueso}},\ }\href {\doibase 10.1038/s41563-025-02261-3}
  {\bibfield  {journal} {\bibinfo  {journal} {Nature Materials}\ }\textbf
  {\bibinfo {volume} {24}},\ \bibinfo {pages} {1005} (\bibinfo {year}
  {2025})}\BibitemShut {NoStop}%
\bibitem [{\citenamefont {Jiang}\ \emph {et~al.}(2025)\citenamefont {Jiang},
  \citenamefont {Holder},\ and\ \citenamefont {Yan}}]{Jiang2025}%
  \BibitemOpen
  \bibfield  {author} {\bibinfo {author} {\bibfnamefont {Y.}~\bibnamefont
  {Jiang}}, \bibinfo {author} {\bibfnamefont {T.}~\bibnamefont {Holder}}, \
  and\ \bibinfo {author} {\bibfnamefont {B.}~\bibnamefont {Yan}},\ }\href
  {\doibase 10.1088/1361-6633/ade454} {\bibfield  {journal} {\bibinfo
  {journal} {Reports on Progress in Physics}\ }\textbf {\bibinfo {volume}
  {88}},\ \bibinfo {pages} {076502} (\bibinfo {year} {2025})}\BibitemShut
  {NoStop}%
\bibitem [{\citenamefont {Fang}\ \emph {et~al.}(2024)\citenamefont {Fang},
  \citenamefont {Cano},\ and\ \citenamefont {Ghorashi}}]{FangYuan2024}%
  \BibitemOpen
  \bibfield  {author} {\bibinfo {author} {\bibfnamefont {Y.}~\bibnamefont
  {Fang}}, \bibinfo {author} {\bibfnamefont {J.}~\bibnamefont {Cano}}, \ and\
  \bibinfo {author} {\bibfnamefont {S.~A.~A.}\ \bibnamefont {Ghorashi}},\
  }\href {\doibase 10.1103/PhysRevLett.133.106701} {\bibfield  {journal}
  {\bibinfo  {journal} {Phys. Rev. Lett.}\ }\textbf {\bibinfo {volume} {133}},\
  \bibinfo {pages} {106701} (\bibinfo {year} {2024})}\BibitemShut {NoStop}%
\bibitem [{\citenamefont {Sala}\ \emph {et~al.}(2025)\citenamefont {Sala},
  \citenamefont {Mercaldo}, \citenamefont {Domi}, \citenamefont {Gariglio},
  \citenamefont {Cuoco}, \citenamefont {Ortix},\ and\ \citenamefont
  {Caviglia}}]{GiacomoSala2025}%
  \BibitemOpen
  \bibfield  {author} {\bibinfo {author} {\bibfnamefont {G.}~\bibnamefont
  {Sala}}, \bibinfo {author} {\bibfnamefont {M.~T.}\ \bibnamefont {Mercaldo}},
  \bibinfo {author} {\bibfnamefont {K.}~\bibnamefont {Domi}}, \bibinfo {author}
  {\bibfnamefont {S.}~\bibnamefont {Gariglio}}, \bibinfo {author}
  {\bibfnamefont {M.}~\bibnamefont {Cuoco}}, \bibinfo {author} {\bibfnamefont
  {C.}~\bibnamefont {Ortix}}, \ and\ \bibinfo {author} {\bibfnamefont {A.~D.}\
  \bibnamefont {Caviglia}},\ }\href {\doibase 10.1126/science.adq3255}
  {\bibfield  {journal} {\bibinfo  {journal} {Science}\ }\textbf {\bibinfo
  {volume} {389}},\ \bibinfo {pages} {822} (\bibinfo {year} {2025})},\ \Eprint
  {http://arxiv.org/abs/https://www.science.org/doi/pdf/10.1126/science.adq3255}
  {https://www.science.org/doi/pdf/10.1126/science.adq3255} \BibitemShut
  {NoStop}%
\bibitem [{\citenamefont {Shao}\ \emph {et~al.}(2020)\citenamefont {Shao},
  \citenamefont {Zhang}, \citenamefont {Gurung}, \citenamefont {Yang},\ and\
  \citenamefont {Tsymbal}}]{ShaoDingFU2020}%
  \BibitemOpen
  \bibfield  {author} {\bibinfo {author} {\bibfnamefont {D.-F.}\ \bibnamefont
  {Shao}}, \bibinfo {author} {\bibfnamefont {S.-H.}\ \bibnamefont {Zhang}},
  \bibinfo {author} {\bibfnamefont {G.}~\bibnamefont {Gurung}}, \bibinfo
  {author} {\bibfnamefont {W.}~\bibnamefont {Yang}}, \ and\ \bibinfo {author}
  {\bibfnamefont {E.~Y.}\ \bibnamefont {Tsymbal}},\ }\href {\doibase
  10.1103/PhysRevLett.124.067203} {\bibfield  {journal} {\bibinfo  {journal}
  {Phys. Rev. Lett.}\ }\textbf {\bibinfo {volume} {124}},\ \bibinfo {pages}
  {067203} (\bibinfo {year} {2020})}\BibitemShut {NoStop}%
\bibitem [{\citenamefont {Zhao}\ \emph {et~al.}(2023)\citenamefont {Zhao},
  \citenamefont {Cao}, \citenamefont {Zhang}, \citenamefont {Li}, \citenamefont
  {Li}, \citenamefont {Ma},\ and\ \citenamefont {Yang}}]{Zhao2023}%
  \BibitemOpen
  \bibfield  {author} {\bibinfo {author} {\bibfnamefont {Y.}~\bibnamefont
  {Zhao}}, \bibinfo {author} {\bibfnamefont {J.}~\bibnamefont {Cao}}, \bibinfo
  {author} {\bibfnamefont {Z.}~\bibnamefont {Zhang}}, \bibinfo {author}
  {\bibfnamefont {S.}~\bibnamefont {Li}}, \bibinfo {author} {\bibfnamefont
  {Y.}~\bibnamefont {Li}}, \bibinfo {author} {\bibfnamefont {F.}~\bibnamefont
  {Ma}}, \ and\ \bibinfo {author} {\bibfnamefont {S.~A.}\ \bibnamefont
  {Yang}},\ }\href {\doibase 10.1103/PhysRevB.107.205124} {\bibfield  {journal}
  {\bibinfo  {journal} {Phys. Rev. B}\ }\textbf {\bibinfo {volume} {107}},\
  \bibinfo {pages} {205124} (\bibinfo {year} {2023})}\BibitemShut {NoStop}%
\bibitem [{\citenamefont {You}\ \emph {et~al.}(2018)\citenamefont {You},
  \citenamefont {Fang}, \citenamefont {Xu}, \citenamefont {Kaxiras},\ and\
  \citenamefont {Low}}]{You2018}%
  \BibitemOpen
  \bibfield  {author} {\bibinfo {author} {\bibfnamefont {J.-S.}\ \bibnamefont
  {You}}, \bibinfo {author} {\bibfnamefont {S.}~\bibnamefont {Fang}}, \bibinfo
  {author} {\bibfnamefont {S.-Y.}\ \bibnamefont {Xu}}, \bibinfo {author}
  {\bibfnamefont {E.}~\bibnamefont {Kaxiras}}, \ and\ \bibinfo {author}
  {\bibfnamefont {T.}~\bibnamefont {Low}},\ }\href {\doibase
  10.1103/PhysRevB.98.121109} {\bibfield  {journal} {\bibinfo  {journal} {Phys.
  Rev. B}\ }\textbf {\bibinfo {volume} {98}},\ \bibinfo {pages} {121109}
  (\bibinfo {year} {2018})}\BibitemShut {NoStop}%
\bibitem [{\citenamefont {Wang}\ \emph {et~al.}(2021)\citenamefont {Wang},
  \citenamefont {Gao},\ and\ \citenamefont {Xiao}}]{WangChong2021}%
  \BibitemOpen
  \bibfield  {author} {\bibinfo {author} {\bibfnamefont {C.}~\bibnamefont
  {Wang}}, \bibinfo {author} {\bibfnamefont {Y.}~\bibnamefont {Gao}}, \ and\
  \bibinfo {author} {\bibfnamefont {D.}~\bibnamefont {Xiao}},\ }\href {\doibase
  10.1103/PhysRevLett.127.277201} {\bibfield  {journal} {\bibinfo  {journal}
  {Phys. Rev. Lett.}\ }\textbf {\bibinfo {volume} {127}},\ \bibinfo {pages}
  {277201} (\bibinfo {year} {2021})}\BibitemShut {NoStop}%
\bibitem [{\citenamefont {Das}\ \emph {et~al.}(2023)\citenamefont {Das},
  \citenamefont {Lahiri}, \citenamefont {Atencia}, \citenamefont {Culcer},\
  and\ \citenamefont {Agarwal}}]{DasKamal2023}%
  \BibitemOpen
  \bibfield  {author} {\bibinfo {author} {\bibfnamefont {K.}~\bibnamefont
  {Das}}, \bibinfo {author} {\bibfnamefont {S.}~\bibnamefont {Lahiri}},
  \bibinfo {author} {\bibfnamefont {R.~B.}\ \bibnamefont {Atencia}}, \bibinfo
  {author} {\bibfnamefont {D.}~\bibnamefont {Culcer}}, \ and\ \bibinfo {author}
  {\bibfnamefont {A.}~\bibnamefont {Agarwal}},\ }\href {\doibase
  10.1103/PhysRevB.108.L201405} {\bibfield  {journal} {\bibinfo  {journal}
  {Phys. Rev. B}\ }\textbf {\bibinfo {volume} {108}},\ \bibinfo {pages}
  {L201405} (\bibinfo {year} {2023})}\BibitemShut {NoStop}%
\bibitem [{\citenamefont {Sharma}\ and\ \citenamefont
  {De~Sarkar}(2025)}]{Sharma2025}%
  \BibitemOpen
  \bibfield  {author} {\bibinfo {author} {\bibfnamefont {S.}~\bibnamefont
  {Sharma}}\ and\ \bibinfo {author} {\bibfnamefont {A.}~\bibnamefont
  {De~Sarkar}},\ }\href {\doibase 10.1103/jy59-jt8s} {\bibfield  {journal}
  {\bibinfo  {journal} {Phys. Rev. B}\ }\textbf {\bibinfo {volume} {112}},\
  \bibinfo {pages} {115435} (\bibinfo {year} {2025})}\BibitemShut {NoStop}%
\bibitem [{\citenamefont {Go}\ \emph {et~al.}(2021)\citenamefont {Go},
  \citenamefont {Jo}, \citenamefont {Lee}, \citenamefont {Kl{\"a}ui},\ and\
  \citenamefont {Mokrousov}}]{Go2021}%
  \BibitemOpen
  \bibfield  {author} {\bibinfo {author} {\bibfnamefont {D.}~\bibnamefont
  {Go}}, \bibinfo {author} {\bibfnamefont {D.}~\bibnamefont {Jo}}, \bibinfo
  {author} {\bibfnamefont {H.-W.}\ \bibnamefont {Lee}}, \bibinfo {author}
  {\bibfnamefont {M.}~\bibnamefont {Kl{\"a}ui}}, \ and\ \bibinfo {author}
  {\bibfnamefont {Y.}~\bibnamefont {Mokrousov}},\ }\href {\doibase
  10.1209/0295-5075/ac2653} {\bibfield  {journal} {\bibinfo  {journal}
  {Europhysics Letters}\ }\textbf {\bibinfo {volume} {135}},\ \bibinfo {pages}
  {37001} (\bibinfo {year} {2021})}\BibitemShut {NoStop}%
\bibitem [{\citenamefont {Atencia}\ \emph {et~al.}(2024)\citenamefont
  {Atencia}, \citenamefont {Agarwal},\ and\ \citenamefont
  {Culcer}}]{BurgosAtencia2024}%
  \BibitemOpen
  \bibfield  {author} {\bibinfo {author} {\bibfnamefont {R.~B.}\ \bibnamefont
  {Atencia}}, \bibinfo {author} {\bibfnamefont {A.}~\bibnamefont {Agarwal}}, \
  and\ \bibinfo {author} {\bibfnamefont {D.}~\bibnamefont {Culcer}},\ }\href
  {\doibase 10.1080/23746149.2024.2371972} {\bibfield  {journal} {\bibinfo
  {journal} {Advances in Physics: X}\ }\textbf {\bibinfo {volume} {9}},\
  \bibinfo {pages} {2371972} (\bibinfo {year} {2024})},\ \Eprint
  {http://arxiv.org/abs/https://doi.org/10.1080/23746149.2024.2371972}
  {https://doi.org/10.1080/23746149.2024.2371972} \BibitemShut {NoStop}%
\bibitem [{\citenamefont {Sundaram}\ and\ \citenamefont
  {Niu}(1999)}]{Sundaram1999}%
  \BibitemOpen
  \bibfield  {author} {\bibinfo {author} {\bibfnamefont {G.}~\bibnamefont
  {Sundaram}}\ and\ \bibinfo {author} {\bibfnamefont {Q.}~\bibnamefont {Niu}},\
  }\href {\doibase 10.1103/PhysRevB.59.14915} {\bibfield  {journal} {\bibinfo
  {journal} {Phys. Rev. B}\ }\textbf {\bibinfo {volume} {59}},\ \bibinfo
  {pages} {14915} (\bibinfo {year} {1999})}\BibitemShut {NoStop}%
\bibitem [{\citenamefont {Chang}\ and\ \citenamefont {Niu}(2008)}]{Chang2008}%
  \BibitemOpen
  \bibfield  {author} {\bibinfo {author} {\bibfnamefont {M.-C.}\ \bibnamefont
  {Chang}}\ and\ \bibinfo {author} {\bibfnamefont {Q.}~\bibnamefont {Niu}},\
  }\href {\doibase 10.1088/0953-8984/20/19/193202} {\bibfield  {journal}
  {\bibinfo  {journal} {Journal of Physics: Condensed Matter}\ }\textbf
  {\bibinfo {volume} {20}},\ \bibinfo {pages} {193202} (\bibinfo {year}
  {2008})}\BibitemShut {NoStop}%
\bibitem [{\citenamefont {Cysne}\ \emph
  {et~al.}(2021{\natexlab{a}})\citenamefont {Cysne}, \citenamefont
  {Guimar\~aes}, \citenamefont {Canonico}, \citenamefont {Rappoport},\ and\
  \citenamefont {Muniz}}]{CysneTarik2021}%
  \BibitemOpen
  \bibfield  {author} {\bibinfo {author} {\bibfnamefont {T.~P.}\ \bibnamefont
  {Cysne}}, \bibinfo {author} {\bibfnamefont {F.~S.~M.}\ \bibnamefont
  {Guimar\~aes}}, \bibinfo {author} {\bibfnamefont {L.~M.}\ \bibnamefont
  {Canonico}}, \bibinfo {author} {\bibfnamefont {T.~G.}\ \bibnamefont
  {Rappoport}}, \ and\ \bibinfo {author} {\bibfnamefont {R.~B.}\ \bibnamefont
  {Muniz}},\ }\href {\doibase 10.1103/PhysRevB.104.165403} {\bibfield
  {journal} {\bibinfo  {journal} {Phys. Rev. B}\ }\textbf {\bibinfo {volume}
  {104}},\ \bibinfo {pages} {165403} (\bibinfo {year}
  {2021}{\natexlab{a}})}\BibitemShut {NoStop}%
\bibitem [{\citenamefont {Cysne}\ \emph {et~al.}(2023)\citenamefont {Cysne},
  \citenamefont {Guimar\~aes}, \citenamefont {Canonico}, \citenamefont {Costa},
  \citenamefont {Rappoport},\ and\ \citenamefont {Muniz}}]{CysneTarik2023}%
  \BibitemOpen
  \bibfield  {author} {\bibinfo {author} {\bibfnamefont {T.~P.}\ \bibnamefont
  {Cysne}}, \bibinfo {author} {\bibfnamefont {F.~S.~M.}\ \bibnamefont
  {Guimar\~aes}}, \bibinfo {author} {\bibfnamefont {L.~M.}\ \bibnamefont
  {Canonico}}, \bibinfo {author} {\bibfnamefont {M.}~\bibnamefont {Costa}},
  \bibinfo {author} {\bibfnamefont {T.~G.}\ \bibnamefont {Rappoport}}, \ and\
  \bibinfo {author} {\bibfnamefont {R.~B.}\ \bibnamefont {Muniz}},\ }\href
  {\doibase 10.1103/PhysRevB.107.115402} {\bibfield  {journal} {\bibinfo
  {journal} {Phys. Rev. B}\ }\textbf {\bibinfo {volume} {107}},\ \bibinfo
  {pages} {115402} (\bibinfo {year} {2023})}\BibitemShut {NoStop}%
\bibitem [{\citenamefont {Lee}\ \emph {et~al.}(2024)\citenamefont {Lee},
  \citenamefont {Park},\ and\ \citenamefont {Lee}}]{LeeJonqiun2024}%
  \BibitemOpen
  \bibfield  {author} {\bibinfo {author} {\bibfnamefont {J.~M.}\ \bibnamefont
  {Lee}}, \bibinfo {author} {\bibfnamefont {M.~J.}\ \bibnamefont {Park}}, \
  and\ \bibinfo {author} {\bibfnamefont {H.-W.}\ \bibnamefont {Lee}},\ }\href
  {\doibase 10.1103/PhysRevB.110.134436} {\bibfield  {journal} {\bibinfo
  {journal} {Phys. Rev. B}\ }\textbf {\bibinfo {volume} {110}},\ \bibinfo
  {pages} {134436} (\bibinfo {year} {2024})}\BibitemShut {NoStop}%
\bibitem [{\citenamefont {Gao}\ \emph {et~al.}(2025)\citenamefont {Gao},
  \citenamefont {Liao}, \citenamefont {Isshiki}, \citenamefont {Budai},
  \citenamefont {Kim}, \citenamefont {Lee}, \citenamefont {Lee}, \citenamefont
  {Go}, \citenamefont {Mokrousov}, \citenamefont {Miwa},\ and\ \citenamefont
  {Otani}}]{Gao2025}%
  \BibitemOpen
  \bibfield  {author} {\bibinfo {author} {\bibfnamefont {W.}~\bibnamefont
  {Gao}}, \bibinfo {author} {\bibfnamefont {L.}~\bibnamefont {Liao}}, \bibinfo
  {author} {\bibfnamefont {H.}~\bibnamefont {Isshiki}}, \bibinfo {author}
  {\bibfnamefont {N.}~\bibnamefont {Budai}}, \bibinfo {author} {\bibfnamefont
  {J.}~\bibnamefont {Kim}}, \bibinfo {author} {\bibfnamefont {H.-W.}\
  \bibnamefont {Lee}}, \bibinfo {author} {\bibfnamefont {K.-J.}\ \bibnamefont
  {Lee}}, \bibinfo {author} {\bibfnamefont {D.}~\bibnamefont {Go}}, \bibinfo
  {author} {\bibfnamefont {Y.}~\bibnamefont {Mokrousov}}, \bibinfo {author}
  {\bibfnamefont {S.}~\bibnamefont {Miwa}}, \ and\ \bibinfo {author}
  {\bibfnamefont {Y.}~\bibnamefont {Otani}},\ }\href {\doibase
  10.1038/s41467-025-61602-7} {\bibfield  {journal} {\bibinfo  {journal}
  {Nature Communications}\ }\textbf {\bibinfo {volume} {16}},\ \bibinfo {pages}
  {6380} (\bibinfo {year} {2025})}\BibitemShut {NoStop}%
\bibitem [{\citenamefont {Canonico}\ \emph
  {et~al.}(2020{\natexlab{a}})\citenamefont {Canonico}, \citenamefont {Cysne},
  \citenamefont {Rappoport},\ and\ \citenamefont {Muniz}}]{Canonico2020}%
  \BibitemOpen
  \bibfield  {author} {\bibinfo {author} {\bibfnamefont {L.~M.}\ \bibnamefont
  {Canonico}}, \bibinfo {author} {\bibfnamefont {T.~P.}\ \bibnamefont {Cysne}},
  \bibinfo {author} {\bibfnamefont {T.~G.}\ \bibnamefont {Rappoport}}, \ and\
  \bibinfo {author} {\bibfnamefont {R.~B.}\ \bibnamefont {Muniz}},\ }\href
  {\doibase 10.1103/PhysRevB.101.075429} {\bibfield  {journal} {\bibinfo
  {journal} {Phys. Rev. B}\ }\textbf {\bibinfo {volume} {101}},\ \bibinfo
  {pages} {075429} (\bibinfo {year} {2020}{\natexlab{a}})}\BibitemShut
  {NoStop}%
\bibitem [{\citenamefont {Canonico}\ \emph
  {et~al.}(2020{\natexlab{b}})\citenamefont {Canonico}, \citenamefont {Cysne},
  \citenamefont {Molina-Sanchez}, \citenamefont {Muniz},\ and\ \citenamefont
  {Rappoport}}]{CanonicoLuis2020}%
  \BibitemOpen
  \bibfield  {author} {\bibinfo {author} {\bibfnamefont {L.~M.}\ \bibnamefont
  {Canonico}}, \bibinfo {author} {\bibfnamefont {T.~P.}\ \bibnamefont {Cysne}},
  \bibinfo {author} {\bibfnamefont {A.}~\bibnamefont {Molina-Sanchez}},
  \bibinfo {author} {\bibfnamefont {R.~B.}\ \bibnamefont {Muniz}}, \ and\
  \bibinfo {author} {\bibfnamefont {T.~G.}\ \bibnamefont {Rappoport}},\ }\href
  {\doibase 10.1103/PhysRevB.101.161409} {\bibfield  {journal} {\bibinfo
  {journal} {Phys. Rev. B}\ }\textbf {\bibinfo {volume} {101}},\ \bibinfo
  {pages} {161409} (\bibinfo {year} {2020}{\natexlab{b}})}\BibitemShut
  {NoStop}%
\bibitem [{\citenamefont {Bhowal}\ and\ \citenamefont
  {Vignale}(2021)}]{Bhowal2021}%
  \BibitemOpen
  \bibfield  {author} {\bibinfo {author} {\bibfnamefont {S.}~\bibnamefont
  {Bhowal}}\ and\ \bibinfo {author} {\bibfnamefont {G.}~\bibnamefont
  {Vignale}},\ }\href {\doibase 10.1103/PhysRevB.103.195309} {\bibfield
  {journal} {\bibinfo  {journal} {Phys. Rev. B}\ }\textbf {\bibinfo {volume}
  {103}},\ \bibinfo {pages} {195309} (\bibinfo {year} {2021})}\BibitemShut
  {NoStop}%
\bibitem [{\citenamefont {Cysne}\ \emph
  {et~al.}(2021{\natexlab{b}})\citenamefont {Cysne}, \citenamefont {Costa},
  \citenamefont {Canonico}, \citenamefont {Nardelli}, \citenamefont {Muniz},\
  and\ \citenamefont {Rappoport}}]{CysneTarik2021II}%
  \BibitemOpen
  \bibfield  {author} {\bibinfo {author} {\bibfnamefont {T.~P.}\ \bibnamefont
  {Cysne}}, \bibinfo {author} {\bibfnamefont {M.}~\bibnamefont {Costa}},
  \bibinfo {author} {\bibfnamefont {L.~M.}\ \bibnamefont {Canonico}}, \bibinfo
  {author} {\bibfnamefont {M.~B.}\ \bibnamefont {Nardelli}}, \bibinfo {author}
  {\bibfnamefont {R.~B.}\ \bibnamefont {Muniz}}, \ and\ \bibinfo {author}
  {\bibfnamefont {T.~G.}\ \bibnamefont {Rappoport}},\ }\href {\doibase
  10.1103/PhysRevLett.126.056601} {\bibfield  {journal} {\bibinfo  {journal}
  {Phys. Rev. Lett.}\ }\textbf {\bibinfo {volume} {126}},\ \bibinfo {pages}
  {056601} (\bibinfo {year} {2021}{\natexlab{b}})}\BibitemShut {NoStop}%
\bibitem [{\citenamefont {Cysne}\ \emph {et~al.}(2022)\citenamefont {Cysne},
  \citenamefont {Bhowal}, \citenamefont {Vignale},\ and\ \citenamefont
  {Rappoport}}]{CysneTarik2022}%
  \BibitemOpen
  \bibfield  {author} {\bibinfo {author} {\bibfnamefont {T.~P.}\ \bibnamefont
  {Cysne}}, \bibinfo {author} {\bibfnamefont {S.}~\bibnamefont {Bhowal}},
  \bibinfo {author} {\bibfnamefont {G.}~\bibnamefont {Vignale}}, \ and\
  \bibinfo {author} {\bibfnamefont {T.~G.}\ \bibnamefont {Rappoport}},\ }\href
  {\doibase 10.1103/PhysRevB.105.195421} {\bibfield  {journal} {\bibinfo
  {journal} {Phys. Rev. B}\ }\textbf {\bibinfo {volume} {105}},\ \bibinfo
  {pages} {195421} (\bibinfo {year} {2022})}\BibitemShut {NoStop}%
\bibitem [{\citenamefont {Pezo}\ \emph {et~al.}(2022)\citenamefont {Pezo},
  \citenamefont {Garc\'{\i}a~Ovalle},\ and\ \citenamefont
  {Manchon}}]{PezoArmando2022}%
  \BibitemOpen
  \bibfield  {author} {\bibinfo {author} {\bibfnamefont {A.}~\bibnamefont
  {Pezo}}, \bibinfo {author} {\bibfnamefont {D.}~\bibnamefont
  {Garc\'{\i}a~Ovalle}}, \ and\ \bibinfo {author} {\bibfnamefont
  {A.}~\bibnamefont {Manchon}},\ }\href {\doibase 10.1103/PhysRevB.106.104414}
  {\bibfield  {journal} {\bibinfo  {journal} {Phys. Rev. B}\ }\textbf {\bibinfo
  {volume} {106}},\ \bibinfo {pages} {104414} (\bibinfo {year}
  {2022})}\BibitemShut {NoStop}%
\bibitem [{\citenamefont {Busch}\ \emph {et~al.}(2023)\citenamefont {Busch},
  \citenamefont {Mertig},\ and\ \citenamefont {G\"obel}}]{Busch2023}%
  \BibitemOpen
  \bibfield  {author} {\bibinfo {author} {\bibfnamefont {O.}~\bibnamefont
  {Busch}}, \bibinfo {author} {\bibfnamefont {I.}~\bibnamefont {Mertig}}, \
  and\ \bibinfo {author} {\bibfnamefont {B.}~\bibnamefont {G\"obel}},\ }\href
  {\doibase 10.1103/PhysRevResearch.5.043052} {\bibfield  {journal} {\bibinfo
  {journal} {Phys. Rev. Res.}\ }\textbf {\bibinfo {volume} {5}},\ \bibinfo
  {pages} {043052} (\bibinfo {year} {2023})}\BibitemShut {NoStop}%
\bibitem [{\citenamefont {Pezo}\ \emph {et~al.}(2023)\citenamefont {Pezo},
  \citenamefont {Garc\'{\i}a~Ovalle},\ and\ \citenamefont
  {Manchon}}]{PezoArmando2023II}%
  \BibitemOpen
  \bibfield  {author} {\bibinfo {author} {\bibfnamefont {A.}~\bibnamefont
  {Pezo}}, \bibinfo {author} {\bibfnamefont {D.}~\bibnamefont
  {Garc\'{\i}a~Ovalle}}, \ and\ \bibinfo {author} {\bibfnamefont
  {A.}~\bibnamefont {Manchon}},\ }\href {\doibase 10.1103/PhysRevB.108.075427}
  {\bibfield  {journal} {\bibinfo  {journal} {Phys. Rev. B}\ }\textbf {\bibinfo
  {volume} {108}},\ \bibinfo {pages} {075427} (\bibinfo {year}
  {2023})}\BibitemShut {NoStop}%
\bibitem [{\citenamefont {Barbosa}\ \emph {et~al.}(2024)\citenamefont
  {Barbosa}, \citenamefont {Canonico}, \citenamefont {Garc\'{\i}a},\ and\
  \citenamefont {Rappoport}}]{Barbosa2024}%
  \BibitemOpen
  \bibfield  {author} {\bibinfo {author} {\bibfnamefont {A.~L.~R.}\
  \bibnamefont {Barbosa}}, \bibinfo {author} {\bibfnamefont {L.~M.}\
  \bibnamefont {Canonico}}, \bibinfo {author} {\bibfnamefont {J.~H.}\
  \bibnamefont {Garc\'{\i}a}}, \ and\ \bibinfo {author} {\bibfnamefont {T.~G.}\
  \bibnamefont {Rappoport}},\ }\href {\doibase 10.1103/PhysRevB.110.085412}
  {\bibfield  {journal} {\bibinfo  {journal} {Phys. Rev. B}\ }\textbf {\bibinfo
  {volume} {110}},\ \bibinfo {pages} {085412} (\bibinfo {year}
  {2024})}\BibitemShut {NoStop}%
\bibitem [{\citenamefont {Cysne}\ \emph {et~al.}(2024)\citenamefont {Cysne},
  \citenamefont {Kort-Kamp},\ and\ \citenamefont {Rappoport}}]{CysneTarik2024}%
  \BibitemOpen
  \bibfield  {author} {\bibinfo {author} {\bibfnamefont {T.~P.}\ \bibnamefont
  {Cysne}}, \bibinfo {author} {\bibfnamefont {W.~J.~M.}\ \bibnamefont
  {Kort-Kamp}}, \ and\ \bibinfo {author} {\bibfnamefont {T.~G.}\ \bibnamefont
  {Rappoport}},\ }\href {\doibase 10.1103/PhysRevResearch.6.023271} {\bibfield
  {journal} {\bibinfo  {journal} {Phys. Rev. Res.}\ }\textbf {\bibinfo {volume}
  {6}},\ \bibinfo {pages} {023271} (\bibinfo {year} {2024})}\BibitemShut
  {NoStop}%
\bibitem [{\citenamefont {Liu}\ and\ \citenamefont
  {Culcer}(2024)}]{LiuHong2024}%
  \BibitemOpen
  \bibfield  {author} {\bibinfo {author} {\bibfnamefont {H.}~\bibnamefont
  {Liu}}\ and\ \bibinfo {author} {\bibfnamefont {D.}~\bibnamefont {Culcer}},\
  }\href {\doibase 10.1103/PhysRevLett.132.186302} {\bibfield  {journal}
  {\bibinfo  {journal} {Phys. Rev. Lett.}\ }\textbf {\bibinfo {volume} {132}},\
  \bibinfo {pages} {186302} (\bibinfo {year} {2024})}\BibitemShut {NoStop}%
\bibitem [{\citenamefont {Robredo}\ \emph {et~al.}(2024)\citenamefont
  {Robredo}, \citenamefont {Schr{\"o}ter}, \citenamefont {Felser},
  \citenamefont {Cano}, \citenamefont {Bradlyn},\ and\ \citenamefont
  {Vergniory}}]{Robredo2024}%
  \BibitemOpen
  \bibfield  {author} {\bibinfo {author} {\bibfnamefont {I.}~\bibnamefont
  {Robredo}}, \bibinfo {author} {\bibfnamefont {N.~B.~M.}\ \bibnamefont
  {Schr{\"o}ter}}, \bibinfo {author} {\bibfnamefont {C.}~\bibnamefont
  {Felser}}, \bibinfo {author} {\bibfnamefont {J.}~\bibnamefont {Cano}},
  \bibinfo {author} {\bibfnamefont {B.}~\bibnamefont {Bradlyn}}, \ and\
  \bibinfo {author} {\bibfnamefont {M.~G.}\ \bibnamefont {Vergniory}},\ }\href
  {\doibase 10.1209/0295-5075/ad6bbc} {\bibfield  {journal} {\bibinfo
  {journal} {Europhysics Letters}\ }\textbf {\bibinfo {volume} {147}},\
  \bibinfo {pages} {46001} (\bibinfo {year} {2024})}\BibitemShut {NoStop}%
\bibitem [{\citenamefont {Ni}\ \emph {et~al.}(2020)\citenamefont {Ni},
  \citenamefont {Xu}, \citenamefont {S{\'a}nchez-Mart{\'\i}nez}, \citenamefont
  {Zhang}, \citenamefont {Manna}, \citenamefont {Bernhard}, \citenamefont
  {Venderbos}, \citenamefont {de~Juan}, \citenamefont {Felser}, \citenamefont
  {Grushin},\ and\ \citenamefont {Wu}}]{NiZhouliang2020}%
  \BibitemOpen
  \bibfield  {author} {\bibinfo {author} {\bibfnamefont {Z.}~\bibnamefont
  {Ni}}, \bibinfo {author} {\bibfnamefont {B.}~\bibnamefont {Xu}}, \bibinfo
  {author} {\bibfnamefont {M.~{\'A}.}\ \bibnamefont
  {S{\'a}nchez-Mart{\'\i}nez}}, \bibinfo {author} {\bibfnamefont
  {Y.}~\bibnamefont {Zhang}}, \bibinfo {author} {\bibfnamefont
  {K.}~\bibnamefont {Manna}}, \bibinfo {author} {\bibfnamefont
  {C.}~\bibnamefont {Bernhard}}, \bibinfo {author} {\bibfnamefont {J.~W.~F.}\
  \bibnamefont {Venderbos}}, \bibinfo {author} {\bibfnamefont {F.}~\bibnamefont
  {de~Juan}}, \bibinfo {author} {\bibfnamefont {C.}~\bibnamefont {Felser}},
  \bibinfo {author} {\bibfnamefont {A.~G.}\ \bibnamefont {Grushin}}, \ and\
  \bibinfo {author} {\bibfnamefont {L.}~\bibnamefont {Wu}},\ }\href {\doibase
  10.1038/s41535-020-00298-y} {\bibfield  {journal} {\bibinfo  {journal} {npj
  Quantum Materials}\ }\textbf {\bibinfo {volume} {5}},\ \bibinfo {pages} {96}
  (\bibinfo {year} {2020})}\BibitemShut {NoStop}%
\bibitem [{\citenamefont {Kaushik}\ and\ \citenamefont
  {Cano}(2021)}]{KaushikSahal2021}%
  \BibitemOpen
  \bibfield  {author} {\bibinfo {author} {\bibfnamefont {S.}~\bibnamefont
  {Kaushik}}\ and\ \bibinfo {author} {\bibfnamefont {J.}~\bibnamefont {Cano}},\
  }\href {\doibase 10.1103/PhysRevB.104.155149} {\bibfield  {journal} {\bibinfo
   {journal} {Phys. Rev. B}\ }\textbf {\bibinfo {volume} {104}},\ \bibinfo
  {pages} {155149} (\bibinfo {year} {2021})}\BibitemShut {NoStop}%
\bibitem [{\citenamefont {Provost}\ and\ \citenamefont
  {Vallee}(1980)}]{Provost1980}%
  \BibitemOpen
  \bibfield  {author} {\bibinfo {author} {\bibfnamefont {J.~P.}\ \bibnamefont
  {Provost}}\ and\ \bibinfo {author} {\bibfnamefont {G.}~\bibnamefont
  {Vallee}},\ }\href {\doibase 10.1007/BF02193559} {\bibfield  {journal}
  {\bibinfo  {journal} {Communications in Mathematical Physics}\ }\textbf
  {\bibinfo {volume} {76}},\ \bibinfo {pages} {289} (\bibinfo {year}
  {1980})}\BibitemShut {NoStop}%
\bibitem [{\citenamefont {Yu}\ \emph {et~al.}(2025)\citenamefont {Yu},
  \citenamefont {Bernevig}, \citenamefont {Queiroz}, \citenamefont {Rossi},
  \citenamefont {T{\"o}rm{\"a}},\ and\ \citenamefont {Yang}}]{YuJiabin2025}%
  \BibitemOpen
  \bibfield  {author} {\bibinfo {author} {\bibfnamefont {J.}~\bibnamefont
  {Yu}}, \bibinfo {author} {\bibfnamefont {B.~A.}\ \bibnamefont {Bernevig}},
  \bibinfo {author} {\bibfnamefont {R.}~\bibnamefont {Queiroz}}, \bibinfo
  {author} {\bibfnamefont {E.}~\bibnamefont {Rossi}}, \bibinfo {author}
  {\bibfnamefont {P.}~\bibnamefont {T{\"o}rm{\"a}}}, \ and\ \bibinfo {author}
  {\bibfnamefont {B.-J.}\ \bibnamefont {Yang}},\ }\href {\doibase
  10.1038/s41535-025-00801-3} {\bibfield  {journal} {\bibinfo  {journal} {npj
  Quantum Materials}\ }\textbf {\bibinfo {volume} {10}},\ \bibinfo {pages}
  {101} (\bibinfo {year} {2025})}\BibitemShut {NoStop}%
\bibitem [{\citenamefont {Zhong}\ \emph {et~al.}(2016)\citenamefont {Zhong},
  \citenamefont {Moore},\ and\ \citenamefont {Souza}}]{ZhongShudan2016}%
  \BibitemOpen
  \bibfield  {author} {\bibinfo {author} {\bibfnamefont {S.}~\bibnamefont
  {Zhong}}, \bibinfo {author} {\bibfnamefont {J.~E.}\ \bibnamefont {Moore}}, \
  and\ \bibinfo {author} {\bibfnamefont {I.}~\bibnamefont {Souza}},\ }\href
  {\doibase 10.1103/PhysRevLett.116.077201} {\bibfield  {journal} {\bibinfo
  {journal} {Phys. Rev. Lett.}\ }\textbf {\bibinfo {volume} {116}},\ \bibinfo
  {pages} {077201} (\bibinfo {year} {2016})}\BibitemShut {NoStop}%
\bibitem [{\citenamefont {Yoda}\ \emph {et~al.}(2018)\citenamefont {Yoda},
  \citenamefont {Yokoyama},\ and\ \citenamefont {Murakami}}]{Yoda2018}%
  \BibitemOpen
  \bibfield  {author} {\bibinfo {author} {\bibfnamefont {T.}~\bibnamefont
  {Yoda}}, \bibinfo {author} {\bibfnamefont {T.}~\bibnamefont {Yokoyama}}, \
  and\ \bibinfo {author} {\bibfnamefont {S.}~\bibnamefont {Murakami}},\ }\href
  {\doibase 10.1021/acs.nanolett.7b04300} {\bibfield  {journal} {\bibinfo
  {journal} {Nano Letters}\ }\textbf {\bibinfo {volume} {18}},\ \bibinfo
  {pages} {916} (\bibinfo {year} {2018})}\BibitemShut {NoStop}%
\bibitem [{\citenamefont {Culcer}\ \emph {et~al.}(2017)\citenamefont {Culcer},
  \citenamefont {Sekine},\ and\ \citenamefont {MacDonald}}]{Culcer2017}%
  \BibitemOpen
  \bibfield  {author} {\bibinfo {author} {\bibfnamefont {D.}~\bibnamefont
  {Culcer}}, \bibinfo {author} {\bibfnamefont {A.}~\bibnamefont {Sekine}}, \
  and\ \bibinfo {author} {\bibfnamefont {A.~H.}\ \bibnamefont {MacDonald}},\
  }\href {\doibase 10.1103/PhysRevB.96.035106} {\bibfield  {journal} {\bibinfo
  {journal} {Phys. Rev. B}\ }\textbf {\bibinfo {volume} {96}},\ \bibinfo
  {pages} {035106} (\bibinfo {year} {2017})}\BibitemShut {NoStop}%
\bibitem [{\citenamefont {Atencia}\ \emph {et~al.}(2022)\citenamefont
  {Atencia}, \citenamefont {Niu},\ and\ \citenamefont {Culcer}}]{Burgos2022}%
  \BibitemOpen
  \bibfield  {author} {\bibinfo {author} {\bibfnamefont {R.~B.}\ \bibnamefont
  {Atencia}}, \bibinfo {author} {\bibfnamefont {Q.}~\bibnamefont {Niu}}, \ and\
  \bibinfo {author} {\bibfnamefont {D.}~\bibnamefont {Culcer}},\ }\href
  {\doibase 10.1103/PhysRevResearch.4.013001} {\bibfield  {journal} {\bibinfo
  {journal} {Phys. Rev. Res.}\ }\textbf {\bibinfo {volume} {4}},\ \bibinfo
  {pages} {013001} (\bibinfo {year} {2022})}\BibitemShut {NoStop}%
\bibitem [{\citenamefont {Flicker}\ \emph {et~al.}(2018)\citenamefont
  {Flicker}, \citenamefont {de~Juan}, \citenamefont {Bradlyn}, \citenamefont
  {Morimoto}, \citenamefont {Vergniory},\ and\ \citenamefont
  {Grushin}}]{Flicker2018}%
  \BibitemOpen
  \bibfield  {author} {\bibinfo {author} {\bibfnamefont {F.}~\bibnamefont
  {Flicker}}, \bibinfo {author} {\bibfnamefont {F.}~\bibnamefont {de~Juan}},
  \bibinfo {author} {\bibfnamefont {B.}~\bibnamefont {Bradlyn}}, \bibinfo
  {author} {\bibfnamefont {T.}~\bibnamefont {Morimoto}}, \bibinfo {author}
  {\bibfnamefont {M.~G.}\ \bibnamefont {Vergniory}}, \ and\ \bibinfo {author}
  {\bibfnamefont {A.~G.}\ \bibnamefont {Grushin}},\ }\href {\doibase
  10.1103/PhysRevB.98.155145} {\bibfield  {journal} {\bibinfo  {journal} {Phys.
  Rev. B}\ }\textbf {\bibinfo {volume} {98}},\ \bibinfo {pages} {155145}
  (\bibinfo {year} {2018})}\BibitemShut {NoStop}%
\bibitem [{\citenamefont {Ye}\ \emph {et~al.}(2018)\citenamefont {Ye},
  \citenamefont {Kang}, \citenamefont {Liu}, \citenamefont {von Cube},
  \citenamefont {Wicker}, \citenamefont {Suzuki}, \citenamefont {Jozwiak},
  \citenamefont {Bostwick}, \citenamefont {Rotenberg}, \citenamefont {Bell},
  \citenamefont {Fu}, \citenamefont {Comin},\ and\ \citenamefont
  {Checkelsky}}]{YeLinda2018}%
  \BibitemOpen
  \bibfield  {author} {\bibinfo {author} {\bibfnamefont {L.}~\bibnamefont
  {Ye}}, \bibinfo {author} {\bibfnamefont {M.}~\bibnamefont {Kang}}, \bibinfo
  {author} {\bibfnamefont {J.}~\bibnamefont {Liu}}, \bibinfo {author}
  {\bibfnamefont {F.}~\bibnamefont {von Cube}}, \bibinfo {author}
  {\bibfnamefont {C.~R.}\ \bibnamefont {Wicker}}, \bibinfo {author}
  {\bibfnamefont {T.}~\bibnamefont {Suzuki}}, \bibinfo {author} {\bibfnamefont
  {C.}~\bibnamefont {Jozwiak}}, \bibinfo {author} {\bibfnamefont
  {A.}~\bibnamefont {Bostwick}}, \bibinfo {author} {\bibfnamefont
  {E.}~\bibnamefont {Rotenberg}}, \bibinfo {author} {\bibfnamefont {D.~C.}\
  \bibnamefont {Bell}}, \bibinfo {author} {\bibfnamefont {L.}~\bibnamefont
  {Fu}}, \bibinfo {author} {\bibfnamefont {R.}~\bibnamefont {Comin}}, \ and\
  \bibinfo {author} {\bibfnamefont {J.~G.}\ \bibnamefont {Checkelsky}},\ }\href
  {\doibase 10.1038/nature25987} {\bibfield  {journal} {\bibinfo  {journal}
  {Nature}\ }\textbf {\bibinfo {volume} {555}},\ \bibinfo {pages} {638}
  (\bibinfo {year} {2018})}\BibitemShut {NoStop}%
\bibitem [{\citenamefont {Kang}\ \emph
  {et~al.}(2020{\natexlab{a}})\citenamefont {Kang}, \citenamefont {Fang},
  \citenamefont {Ye}, \citenamefont {Po}, \citenamefont {Denlinger},
  \citenamefont {Jozwiak}, \citenamefont {Bostwick}, \citenamefont {Rotenberg},
  \citenamefont {Kaxiras}, \citenamefont {Checkelsky},\ and\ \citenamefont
  {Comin}}]{KangMinguII2020}%
  \BibitemOpen
  \bibfield  {author} {\bibinfo {author} {\bibfnamefont {M.}~\bibnamefont
  {Kang}}, \bibinfo {author} {\bibfnamefont {S.}~\bibnamefont {Fang}}, \bibinfo
  {author} {\bibfnamefont {L.}~\bibnamefont {Ye}}, \bibinfo {author}
  {\bibfnamefont {H.~C.}\ \bibnamefont {Po}}, \bibinfo {author} {\bibfnamefont
  {J.}~\bibnamefont {Denlinger}}, \bibinfo {author} {\bibfnamefont
  {C.}~\bibnamefont {Jozwiak}}, \bibinfo {author} {\bibfnamefont
  {A.}~\bibnamefont {Bostwick}}, \bibinfo {author} {\bibfnamefont
  {E.}~\bibnamefont {Rotenberg}}, \bibinfo {author} {\bibfnamefont
  {E.}~\bibnamefont {Kaxiras}}, \bibinfo {author} {\bibfnamefont {J.~G.}\
  \bibnamefont {Checkelsky}}, \ and\ \bibinfo {author} {\bibfnamefont
  {R.}~\bibnamefont {Comin}},\ }\href {\doibase 10.1038/s41467-020-17465-1}
  {\bibfield  {journal} {\bibinfo  {journal} {Nature Communications}\ }\textbf
  {\bibinfo {volume} {11}},\ \bibinfo {pages} {4004} (\bibinfo {year}
  {2020}{\natexlab{a}})}\BibitemShut {NoStop}%
\bibitem [{\citenamefont {Kang}\ \emph
  {et~al.}(2020{\natexlab{b}})\citenamefont {Kang}, \citenamefont {Ye},
  \citenamefont {Fang}, \citenamefont {You}, \citenamefont {Levitan},
  \citenamefont {Han}, \citenamefont {Facio}, \citenamefont {Jozwiak},
  \citenamefont {Bostwick}, \citenamefont {Rotenberg}, \citenamefont {Chan},
  \citenamefont {McDonald}, \citenamefont {Graf}, \citenamefont {Kaznatcheev},
  \citenamefont {Vescovo}, \citenamefont {Bell}, \citenamefont {Kaxiras},
  \citenamefont {van~den Brink}, \citenamefont {Richter}, \citenamefont
  {Prasad~Ghimire}, \citenamefont {Checkelsky},\ and\ \citenamefont
  {Comin}}]{KangMingu2020}%
  \BibitemOpen
  \bibfield  {author} {\bibinfo {author} {\bibfnamefont {M.}~\bibnamefont
  {Kang}}, \bibinfo {author} {\bibfnamefont {L.}~\bibnamefont {Ye}}, \bibinfo
  {author} {\bibfnamefont {S.}~\bibnamefont {Fang}}, \bibinfo {author}
  {\bibfnamefont {J.-S.}\ \bibnamefont {You}}, \bibinfo {author} {\bibfnamefont
  {A.}~\bibnamefont {Levitan}}, \bibinfo {author} {\bibfnamefont
  {M.}~\bibnamefont {Han}}, \bibinfo {author} {\bibfnamefont {J.~I.}\
  \bibnamefont {Facio}}, \bibinfo {author} {\bibfnamefont {C.}~\bibnamefont
  {Jozwiak}}, \bibinfo {author} {\bibfnamefont {A.}~\bibnamefont {Bostwick}},
  \bibinfo {author} {\bibfnamefont {E.}~\bibnamefont {Rotenberg}}, \bibinfo
  {author} {\bibfnamefont {M.~K.}\ \bibnamefont {Chan}}, \bibinfo {author}
  {\bibfnamefont {R.~D.}\ \bibnamefont {McDonald}}, \bibinfo {author}
  {\bibfnamefont {D.}~\bibnamefont {Graf}}, \bibinfo {author} {\bibfnamefont
  {K.}~\bibnamefont {Kaznatcheev}}, \bibinfo {author} {\bibfnamefont
  {E.}~\bibnamefont {Vescovo}}, \bibinfo {author} {\bibfnamefont {D.~C.}\
  \bibnamefont {Bell}}, \bibinfo {author} {\bibfnamefont {E.}~\bibnamefont
  {Kaxiras}}, \bibinfo {author} {\bibfnamefont {J.}~\bibnamefont {van~den
  Brink}}, \bibinfo {author} {\bibfnamefont {M.}~\bibnamefont {Richter}},
  \bibinfo {author} {\bibfnamefont {M.}~\bibnamefont {Prasad~Ghimire}},
  \bibinfo {author} {\bibfnamefont {J.~G.}\ \bibnamefont {Checkelsky}}, \ and\
  \bibinfo {author} {\bibfnamefont {R.}~\bibnamefont {Comin}},\ }\href
  {\doibase 10.1038/s41563-019-0531-0} {\bibfield  {journal} {\bibinfo
  {journal} {Nature Materials}\ }\textbf {\bibinfo {volume} {19}},\ \bibinfo
  {pages} {163} (\bibinfo {year} {2020}{\natexlab{b}})}\BibitemShut {NoStop}%
\bibitem [{\citenamefont {Green}\ \emph {et~al.}(2010)\citenamefont {Green},
  \citenamefont {Santos},\ and\ \citenamefont {Chamon}}]{GreenDmitry2010}%
  \BibitemOpen
  \bibfield  {author} {\bibinfo {author} {\bibfnamefont {D.}~\bibnamefont
  {Green}}, \bibinfo {author} {\bibfnamefont {L.}~\bibnamefont {Santos}}, \
  and\ \bibinfo {author} {\bibfnamefont {C.}~\bibnamefont {Chamon}},\ }\href
  {\doibase 10.1103/PhysRevB.82.075104} {\bibfield  {journal} {\bibinfo
  {journal} {Phys. Rev. B}\ }\textbf {\bibinfo {volume} {82}},\ \bibinfo
  {pages} {075104} (\bibinfo {year} {2010})}\BibitemShut {NoStop}%
\bibitem [{\citenamefont {Yang}\ \emph {et~al.}(2023)\citenamefont {Yang},
  \citenamefont {Xiao}, \citenamefont {Robredo}, \citenamefont {Vergniory},
  \citenamefont {Yan},\ and\ \citenamefont {Felser}}]{QunYang2023}%
  \BibitemOpen
  \bibfield  {author} {\bibinfo {author} {\bibfnamefont {Q.}~\bibnamefont
  {Yang}}, \bibinfo {author} {\bibfnamefont {J.}~\bibnamefont {Xiao}}, \bibinfo
  {author} {\bibfnamefont {I.}~\bibnamefont {Robredo}}, \bibinfo {author}
  {\bibfnamefont {M.~G.}\ \bibnamefont {Vergniory}}, \bibinfo {author}
  {\bibfnamefont {B.}~\bibnamefont {Yan}}, \ and\ \bibinfo {author}
  {\bibfnamefont {C.}~\bibnamefont {Felser}},\ }\href {\doibase
  10.1073/pnas.2305541120} {\bibfield  {journal} {\bibinfo  {journal}
  {Proceedings of the National Academy of Sciences}\ }\textbf {\bibinfo
  {volume} {120}},\ \bibinfo {pages} {e2305541120} (\bibinfo {year} {2023})},\
  \Eprint
  {http://arxiv.org/abs/https://www.pnas.org/doi/pdf/10.1073/pnas.2305541120}
  {https://www.pnas.org/doi/pdf/10.1073/pnas.2305541120} \BibitemShut {NoStop}%
\end{thebibliography}%

\end{document}